\documentclass[aps,prb,twocolumn,showpacs]{revtex4}
  \usepackage[dvips]{graphicx}

\usepackage{amssymb,amsmath,amsfonts,hyperref}
\usepackage{wasysym,hyperref}
\usepackage{latexsym}
\usepackage{subfigure}

\newcommand{\be}{\begin{equation}}
\newcommand{\ee}{\end{equation}}

\begin{document}

\title{Band touching from real space topology in frustrated hopping models} 
\date{\today}

\author{Doron L. Bergman$^1$, Congjun Wu$^2$,
and Leon Balents$^3$}
\affiliation{
${}^1$Department of Physics, Yale University,
  New Haven, CT 
06520-8120\\
${}^2$Department of Physics, University of California, San Diego, 
La Jolla, CA 92093-0319\\ 
${}^3$Department of Physics, University of California,
  Santa Barbara, CA 
93106-9530
}

\begin{abstract}
  We study ``frustrated'' hopping models, in which at least one energy
  band, at the maximum or minimum of the spectrum, is dispersionless.
  The states of the flat band(s) can be represented in a basis which is
  fully localized, having support on a vanishing fraction of the system
  in the thermodynamic limit.  In the majority of examples, a dispersive
  band touches the flat band(s) at a number of discrete points in
  momentum space.  We demonstrate that this band touching is related to
  states which exhibit non-trivial topology in real space.
  Specifically, these states have support on one-dimensional loops which
  wind around the entire system (with periodic boundary conditions).  A
  counting argument is given that determines, in each case, whether
  there is band touching or not, in precise correspondence to the result
  of straightforward diagonalization.  When they are present, the topological structure
  protects the band touchings in the sense that they can only be removed
  by perturbations which {\sl also} split the degeneracy of the flat
  band.  
\end{abstract}
\date{\today}

\maketitle

\section{introduction}\label{intro}

The theory of ``accidental'' touching of energy bands in crystals has
been recognized and studied since the early days of the quantum theory
of solids.\cite{PhysRev.52.365} By accidental, one means that the
touching is not required by symmetry.  A spectacular example of current
interest is the Dirac point degeneracy of graphene, which leads to a
host of interesting behavior.\cite{geim2007rg} Another class of heavily
studied theoretical examples are the Dirac points appearing in problems
of two dimensional electrons moving in periodic potentials in a magnetic
field studied by Hofstadter\cite{hofstadter1976ela} and others. Three
dimensional Dirac points occur in models of unusual ``spin Hall
insulators'' occurring with strong spin-orbit
interactions.\cite{Murakami:2007} In all these cases, despite the
accidental nature of the band touching, it is robust to perturbations of
the Hamiltonian.  This robustness has its origin in {\sl momentum space
  topology} of the Bloch wavefunctions.\cite{Manes:prb07} For instance,
in graphene, each Dirac point is a source of a $\pm \pi$ delta-function
flux of Berry curvature, so that the line integral of the Berry
connection, $\oint_{\cal C} d\vec{k}\cdot {\rm Im} \langle
u_{nk}|\vec\nabla_{k}|u_{nk}\rangle = \pm \pi$ for any curve ${\cal C}$
enclosing a Dirac point.  If time reversal and inversion symmetries are
maintained, the Berry curvature vanishes identically except at points of
band crossing, and conservation of its flux protects the band
crossings comprising the Dirac points. 

In this paper, we describe the topological protection behind a
completely different instance of accidental band touching, which
occurs in a broad class of ``frustrated'' hopping models.  The models
which we will consider actually display in addition to band crossings a
more dramatic phenomena: the presence of one or more {\sl completely
  flat bands}.  Models with flat bands are particularly interesting
physically because in this case the effect of interactions is wholly
non-perturbative: interactions can reconstruct the states within the
flat band manifold without any cost in kinetic energy.  This is a
powerful mechanism for generating complex and interesting many-body
states, as attested by the richness of the fractional quantum Hall
effect, which occurs as a result of the flat band degeneracy of Landau
levels of electrons in a magnetic field.  

The frustrated hopping models we consider here arise in other contexts,
e.g.  the description of magnons in frustrated quantum antiferromagnets,
and the motion of cold atoms in p-wave Bloch bands optical lattices.
When the density of particles described by such frustrated hopping
models is sufficiently low, short-range interactions of arbitrary (weak
or strong) strength lead to particle localization into a variety of
crystalline patterns that are model
specific.\cite{Wu:prl07,Schulenberg:prl02,Derzhko:prb04,Zhitomirsky:05,Zhitomirsky:06,Derzhko:07,Moessner:2006}
These states are analogous to the Wigner crystal states of electrons in
the Lowest Landau Level (LLL), which indeed occur for sufficiently small
filling factor.  These states can be understood as follows.  From the
flat band, one can construct single particle Wannier states
(superpositions of wavefunctions with all momenta) which are strictly
localized, i.e. have support on only a small finite number of sites.
While Wannier states may always be constructed, only for the case of a
flat band do they remain one-particle eigenstates.  When interactions
also have finite range and are repulsive, multiple particles can be
present in spatially separated localized states with zero interaction
cost, and the corresponding many-body wavefunction remains a many-body
eigenstate.  If the flat band has the minimum kinetic energy, such a
state minimizes simultaneously kinetic and interaction energy, and is
therefore a ground state.  Such states are generically possible for
particle densities below some ``close packing'' threshold, at which
interaction cost becomes inevitable.  The states precisely at this
threshold density are usually (but not always) periodic crystalline
configurations.

In the case of electrons in the LLL, the more interesting fractional
quantum Hall states occur {\sl above} this filling factor, when there is
some unavoidable interaction cost.  However, because of the presence of
a (large) gap between the LLL and the first LL, the interactions act
entirely within the former.  The zoo of fractional quantum Hall states
is understood primarily from studies of the Coulomb interaction {\sl
  projected} into the LLL subspace.  A very interesting question is
whether any similar richness of behavior might occur, above the close
packing threshold, when interactions are included in the frustrated
hopping models discussed here.  This would appear a promising place to
search for exotic orderless quantum spin liquid phases, loosely
analogous to fractional quantum Hall liquids, that have been
hypothesized to occur e.g. in frustrated
magnets\cite{Anderson:RVB,Fazekas:RVB}.

At this point the band touchings re-enter the picture as a {\sl
  hindrance} to this search. The projection of the Hamiltonian into the
lowest energy flat band is strictly controlled only when there is a gap
between this band and the higher dispersive ones, and when this gap is
large compared to the strength of interactions.  In the vast majority of
frustrated hopping models (we will catalog many below), the gap actually
vanishes due to touchings of the first excited band with the flat one at
specific points in momentum space.  Before attempting to surmount this
obstacle, it is crucial to know in each case whether this touching can
removed by some small change in the Hamiltonian, or whether it somehow
enjoys protection that makes a search for such perturbations fruitless.
The result of this paper is that in many cases the crossing {\sl is}
protected, and can only be removed by perturbations that also destroy
the flatness of the low energy band.  Like the protection of the Dirac
points of graphene and others discussed above, the mechanism for this
stability is topological.  However, because of the localized character
of the states in the flat band, the topological structure lies in {\sl
  real space} rather than momentum space.  Specifically, the band
touchings can be associated with eigenstates whose support is extended
along {\sl non-contractible loops} crossing a (toroidal) sample with
periodic boundary conditions.  

The remainder of this paper is organized as follows.  \
In Sec.~\ref{sec:kagome-lattice-model}, we
describe in detail the structure of local and topological loop states
for one of our simplest examples, the nearest-neighbor hopping model on
a kagome lattice.  We show how counting of these states requires band
touching.  In Sec.~\ref{local_eigenstates in various models}, we give
a more abbreviated presentation of the generalization of these arguments
to various other frustrated lattices.  Finally, we conclude with a
discussion in Sec.~\ref{discussion}.  An appendix (\ref{appendix1}) describes an
additional model with further-neighbor interactions without band
touchings. 

\section{Kagome lattice model}
\label{sec:kagome-lattice-model}

The simplest  model we will consider is the 
nearest neighbor tight hopping Hamiltonian on the kagome lattice, 
\begin{equation}\label{hopping}
{\mathcal H}_t = - t \sum_{\langle i j \rangle} \left( c^{\dagger}_i c^{\phantom \dagger}_j + h.c. \right) 
\; ,
\end{equation}
where the indices $i,j$ denote the sites of the kagome lattice,
${\langle i j \rangle}$ denotes nearest neighbor pairs of sites, and the
particles can be either Fermions or Bosons. We use this simple model to
demonstrate all the generic flat band features discussed above.  Aside
of providing concrete examples for all the unique features of flat
bands, the analysis of the kagome model also proves a good model with
which to develop the techniques we will use throughout this manuscript.

\subsection{Band structure}
\label{sec:band-structure}

\begin{figure}
	\centering
		\includegraphics[width=2.0in]{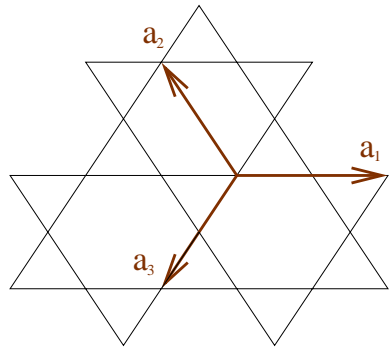}
	\caption{(Color online) Conventions for the shortest length Bravais lattice vectors for the kagome lattice (brown arrows).}
	\label{fig:convention}
\end{figure}

The band structure of Eq.~(\ref{hopping}) consists of a single flat band
with energy
$\epsilon_0({\bf q}) = 2 t$ and two dispersive bands with
\begin{equation}
\epsilon_{\pm}({\bf q}) = - t \left( 1 \pm \sqrt{3 + 2 \Lambda({\bf q})}\right)
\end{equation}
where $ \Lambda({\bf q}) = \cos({\bf q} \cdot {\bf a}_1) + \cos({\bf q}
\cdot {\bf a}_2) + \cos({\bf q} \cdot {\bf a}_3).  $ Here ${\bf
  a}_{1,2,3}$ are the three shortest Bravais lattice vectors for the
kagome (and triangular) lattice. Our conventions are described
succinctly in Fig.~\ref{fig:convention}.  The upper dispersive band
$\epsilon_{-}({\bf q})$ touches the flat band at the $\Gamma$ point
${\bf q} = 0$.

The hopping term in the band basis is of the form
\begin{equation} {\mathcal H}_t = \sum_{\nu} \int_{\bf q}
  a^{\dagger}_{\nu}({\bf q}) a^{\phantom \dagger}_{\nu}({\bf q})
  \epsilon_{\nu}({\bf q}) \; ,
\end{equation}
where the momentum integration is over the first Brillouin zone, and we
use $\nu = 0,\pm$ for the band index, and $\mu = 1,2,3$ for the basis
index.  The new operators $a^{\phantom \dagger}_{\nu}({\bf q})$
are related to the original operators by a Unitary transformation.
In particular, momentum eigenstates of the flat band consist of 
\begin{equation}\label{eq:2}
a_0({\bf q}) = \sum_{\mu=1}^3 \psi^*_{\mu}({\bf q}) c_{\mu} ({\bf q})
\end{equation}
with 
$
\psi_{\mu}({\bf q}) = \frac{ \sin({\bf q} \cdot {\bf a}_{\mu+2}/2) }{ \sqrt{\left( 3 - \Lambda({\bf q}) \right)/2} }
$, where the greek index arithmetic is always modulo 3, and $\frac{3 - \Lambda({\bf q})}{2} = \sum_{\mu=1}^3 \sin^2({\bf q} \cdot {\bf a}_{\mu+2}/2)$. 

\subsection{Localized states}
\label{sec:localized-states}

\begin{figure}
	\centering
		\includegraphics[width=3.0in]{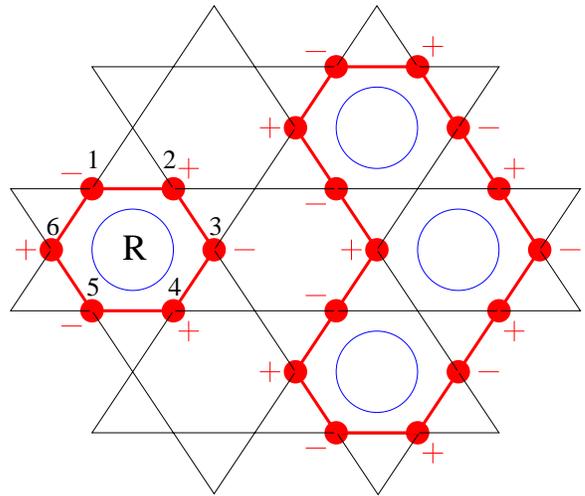}
	\caption{(Color online) Depiction of localized eigenstates, on the boundary of a single and triple plaquette. Those sites with nonzero weight are denoted by a full (red) circle. The magnitude of the weights is always the same, but the phases alternate between $\pm 1$. The phases are denoted by $\pm$ signs next to the relevant lattice sites.}
	\label{fig:kagome_1}
\end{figure}

We can construct localized eigenstates by taking the linear combinations
\begin{equation}
A^{\dagger}_{\bf R} = {\mathcal N}\int_{\bf q} e^{-i {\bf q} \cdot {\bf R} }
a^{\dagger}_0({\bf q}) \sqrt{\left( 3 - \Lambda({\bf q}) \right)/2}
\; ,
\end{equation}
with ${\mathcal N}$ being some normalization.
Here and elsewhere we will use $A^{\dagger}_{\bf R}$ to denote the creation 
operator for the localized eigenstates.
Choosing ${\bf R}$ to be the position at the center of an hexagonal plaquette of the
lattice, and normalizing the operator we find
\begin{equation}\label{plaquette_op}
A^{\dagger}_{\bf R} = \frac{1}{\sqrt{6}} \sum_{j=1}^6 (-1)^j c_j^{\dagger}
\; ,
\end{equation}
where the indices $1 \ldots 6$ enumerate the 6 successive sites around the hexagonal plaquette,
as illustrated in Fig.~\ref{fig:kagome_1}. These local operators
are very useful, but they are unfortunately \emph{not} canonical bosons or fermions. Rather, if $c_j$ are bosonic, the commutation relations are
\begin{equation}
\left[ A^{\phantom\dagger}_{\bf R} , A^{\dagger}_{\bf R'} \right] = 
\delta_{{\bf R}, {\bf R'}} - \frac{1}{6} \Gamma_{{\bf R}, {\bf R'}}
\; ,\label{eq:1}
\end{equation}
where the matrix $\Gamma_{{\bf R}, {\bf R'}}$ is the adjacency matrix of
the \emph{triangular} lattice formed by the centers of the plaquettes.
For fermions, Eq.~(\ref{eq:1}) holds with the commutator replaced by an
anticommutator. 

\begin{figure}
	\centering
		\includegraphics[width=2.0in]{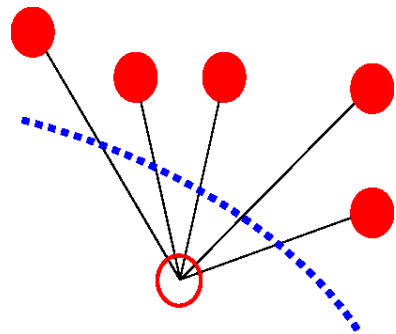}
	\caption{(Color online) The localized states are exact eigenstates due to destructive interference between
	the hopping amplitudes from sites with nonzero weight (filled circles) to sites outside 
	the boundary (empty circle). The lattice sites with nonzero weight are contained in a finite area,
	within a boundary marked by the dashed line.}
	\label{fig:boundary_pic}
\end{figure}

The localized model can be understood directly in real space by
considering a single triangle around the boundary of the plaquette. One
of the corners has an amplitude of $\frac{1}{\sqrt{6}}$, a second has
$\frac{-1}{\sqrt{6}}$ and a third has $0$ amplitude.  The hopping
amplitude from the first and second sites onto the third site
\emph{cancels} out.  Thus the eigenstate is localized as a result of  \emph{destructive
  interference}, which is a very useful guiding principle in
identifying these states in other flat band models.  For a strictly localized
wavefunction to be an eigenstate, the sum of
hopping amplitudes onto sites outside the support of the wavefunction
must vanish (see for illustration
Fig.~\ref{fig:boundary_pic}).

One can create similar exact single-particle eigenstates on larger
loops, by summing over the plaquette states on a number of contiguous
plaquettes, and normalizing the state by the length of the boundary of
the area covered by the plaquettes
\begin{equation}\label{loop_state}
  A^{\dagger}_{\partial{\mathcal A}} = \sum_{\bf R \in {\mathcal A}} A^{\dagger}_{\bf R} \frac{\sqrt{6}}{\sqrt{|\partial{\mathcal A}|}}
  \; .
\end{equation}
Here ${\mathcal{A}}$ denotes the area covered by the plaquettes, and $|\partial{\mathcal A}|$ denotes the 
length of the boundary of this area.
In Fig.~\ref{fig:kagome_1} we show one example of a three-plaquette loop.

\subsection{State counting and band touching}
\label{sec:state-counting}

We now turn to the main question addressed in this paper, of the origin
of the band touching.  We will show that the set of localized
eigenstates contains too many states to fit into the flat band alone.
Specifically, the dimension of the space of localized state with the
{\sl energy} of the flat band has a dimension which is $1$ larger than
that of the flat band.  This requires a contribution from a state of
another band, which, since it is continuous, must touch the flat band at
one point.

Because the difference in question involves only a finite number of
basis states (here $1$, but there may be more in other examples in the
next section), it is necessary to consider a large but finite system to
make this counting precise.  It is advantageous to use periodic boundary
conditions (with a finite integral number of unit cells in each of two
directions), since in this case the Bloch states in Eq.~(\ref{eq:2})
remain eigenstates (with discrete $q$) in the finite system.  We must
count carefully the number of linearly independent states with energy
$\epsilon_0$.  The plaquette states created by Eq.~\eqref{plaquette_op},
na\"ively all seem linearly independent, since they occur on different
plaquettes. With open boundary conditions, the sum over all the
plaquettes in the lattice leads to a state of the form of
Eq.~\eqref{loop_state} at the boundary of the system. For periodic
boundary conditions (putting the lattice on a torus), however, this sum
\emph{vanishes} since there is no boundary $A^{\dagger}_{{\bf q} = 0} =
\sum_{\bf R} A^{\dagger}_{\bf R} = 0$.  So when considering the Hilbert
space spanned by the plaquette states~\eqref{plaquette_op} we have only
$(N-1)$ independent states, where $N$ is the number of plaquettes (and
unit cells) in the lattice. This accounts for all but one state of the
flat band.

\begin{figure}
	\centering
		\includegraphics[width=3.0in]{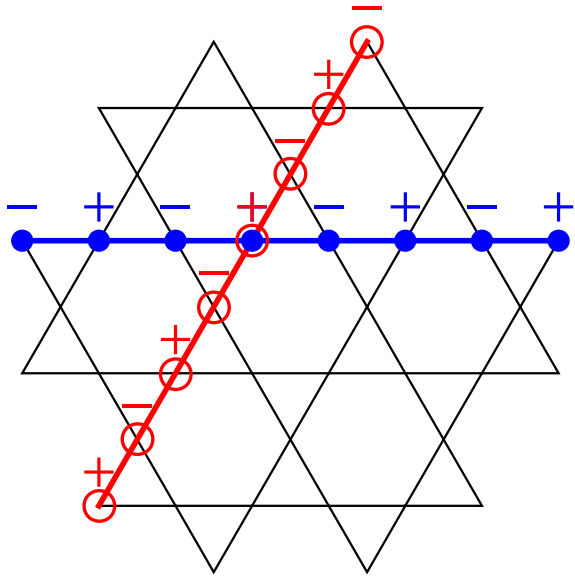}
	\caption{(Color online) The two non-contractible loop states around the handles of the torus.
	One loop consists of the sites marked by full (blue) circles, and the other by the empty (red) circles.
	As the other eigenstates in the flat band, the wavefunction has an alternating $\pm$ phase 
	on the sites along the loops.}
	\label{fig:kagome_2}
\end{figure}

The missing state is accounted for by a non-contractible loop around the
torus. By decorating such a loop with alternating plus/minus signs, as
illustrated in Fig.~\ref{fig:kagome_2}, one again satisfies the
conditions for destructive interference of outgoing waves, and the
associated wavefunction represent an exact eigenstate, with the flat
band energy. This state cannot be expressed as a sum of plaquette
operators, or it would be possible to contract the loop just as any sum
of plaquette states is.  We have therefore found the missing state!
However, we have an embarrassment of riches -- there is not one such
non-contractible loop, but \emph{two}. In total we have $(N+1)$ states,
all with the \emph{same} energy.  From the band structure we know the
flat band contains precisely $N$ states, and so the additional state
must come from another band, and for this reason one of the dispersive
bands touches the flat band at exactly one point.

In fact, from the loop states we can construct the plane wave Bloch
state which touches the flat band explicitly.  By taking an equal weight
linear superposition of the non-contractible loops translated in any
direction other than that along which the loop runs, one obtains a state
with the same configuration in any unit cell, which therefore has the
Bloch form with momentum ${\bf q} = 0$.  The double degeneracy
of states with ${\bf q} = 0$ signifies that not only must one of the
dispersing bands touch the flat band at a point, but that the point is
at ${\bf q} = 0$.

\section{Local eigenstates}\label{local_eigenstates in various models}
\subsection{Pyrochlore lattice model}
\label{sec:pyrochl-latt-model}

Taking the nearest neighbor hopping model \eqref{hopping} on the pyrochlore 
lattice (instead of the kagome lattice) has two degenerate flat bands at $\epsilon_0 = 2t$,
and two dispersive bands $\epsilon_{\pm} = -2t\left( 1 \pm 
\sqrt{1 + \cos{\frac{q_1}{2}} \cos{\frac{q_2}{2}} + \cos{\frac{q_2}{2}} \cos{\frac{q_3}{2}}
 + \cos{\frac{q_3}{2}} \cos{\frac{q_1}{2}}  } 
\right)$, where we have used the conventions ${\bf a}_1 = \frac{1}{2}(0,1,1)$,
${\bf a}_2 = \frac{1}{2}(0,1,1)$, and ${\bf a}_3 = \frac{1}{2}(0,1,1)$ for the (FCC)
Bravais lattice vectors, and ${\bf e}_0 = \frac{1}{8}(1,1,1)$,
${\bf e}_1 = \frac{1}{8}(-1,1,1)$, ${\bf e}_2 = \frac{1}{8}(1,-1,1)$
and ${\bf e}_3 = \frac{1}{8}(1,1,-1)$ for the pyrochlore basis.
Both flat bands touch the upper dispersive band at ${\bf q} = 0$. 
The \emph{same} localized plaquette modes that
appear in the kagome model, are exact eigenstates for this pyrochlore model as well. However,
whereas the number of hexagonal plaquettes in the kagome lattice is equal the number of unit cells, in the 
pyrochlore lattice the number of plaquettes is 4 times that of the number of unit cells. With two flat bands
containing only $2N$ states, clearly these are not all linearly independent.

\begin{figure}
	\centering
		\includegraphics[width=2.5in]{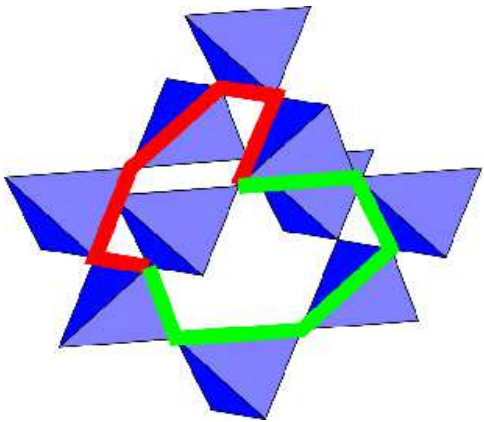}
	\caption{(Color online) Pyrochlore volume enclosed by 4 plaquettes. Two of the 4 plaquettes are highlighted by thick (red and green) lines. Each one of the 2 plaquettes supports a localized eigenstate.}
	\label{fig:pyro}
\end{figure}

Consider a volume enclosed by 4 plaquettes (see Fig~\ref{fig:pyro}). Placing plaquette states with equal weight, 
and appropriate relative signs, on each one of these 4 faces gives a total of \emph{zero}. There are $2N$ such cells in the 
pyrochlore lattice, and therefore $2N$ such constraints. This reduces the number of independent states we can 
construct out of the plaquette states to $2N$. We choose to keep all the plaquette states for the plaquettes 
perpendicular to two out of the 4 $\langle 111 \rangle$ directions of the pyrochlore lattice.

Now if we consider any one of the kagome planes along the two directions we chose above,
we have the same additional constraint as in the kagome lattice - putting a plaquette state on every 
plaquette in the plane, with periodic boundary conditions results in \emph{zero}, giving us one 
additional constraint.
Taking into account the cell-constraints from the previous paragraph, there are only \emph{two} such linearly independent
planes, so we have two additional constraints, reducing the number of linearly independent states we can construct from
the plaquette states to $2N - 2$. As in the kagome lattice, we now have the non-contractible loop states to consider.
In the pyrochlore lattice there will be 3 such non-contractible loops rather than 2. Therefore, in total we 
have $2N +1$ states with the same energy of the flat bands. Exactly as argued in the kagome case, these particular 
non-contractible loop states can be made into ${\bf q} = 0$ states, and we therefore have 3 degenerate states
at ${\bf q} = 0$. The only way for the band structure to comply with this is to have one of the dispersive bands touch 
the two flat bands at the ${\bf q} = 0$ point which indeed is the case.

\subsection{Dice lattice model}
\label{sec:dice-lattice-model}

The nearest neighbor hopping model can give rise to a flat band on
other lattices as well. We considered one example of a 2D
lattice, and one example of a 3D lattice. In this subsection we shall mention one 
additional 2D lattice - the dice lattice, for a number of reasons. First, as opposed to the kagome and 
pyrochlore lattices, in this model the flat band touches dispersive bands at momenta
other than ${\bf q} = 0$. Second, it will be useful to compare between two different 
hopping models on this lattice, that both produce a flat band. The analysis essentially follows the same
steps as in the kagome lattice model, and so we will not elaborate how the results were obtained.

The dice lattice has a basis of 3 sites, two of which have coordination number 3, and one with 
coordination number 6. In what follows, we shall refer to the latter sites as the coordination-6 sites.
On the dice lattice, the nearest neighbor hopping model has one flat band at 
$\epsilon = 0$ touching two dispersive bands 
$\epsilon_{\pm} = \pm 2 t {\sqrt 2} \sqrt{3 + 2  \Lambda({\bf q}) }$ (with $\Lambda({\bf q})$ the same as defined in the introduction) at the two momentum points 
${\bf q} = \pm (\frac{4\pi}{\sqrt 3},0)$ (with the Bravais lattice vectors taken with length 1).
Adding an on-site potential which does not break the symmetries of the lattice(for instance an energy cost $V$ to be on a 6-coordination site), 
one can gap one of the two dispersive bands away from the flat band, and only two degenerate points will remain. Therefore, our counting arguments
will have accounted for $N+2$ states with the energy of the flat band.

\begin{figure}
	\centering
		\includegraphics[width=3.0in]{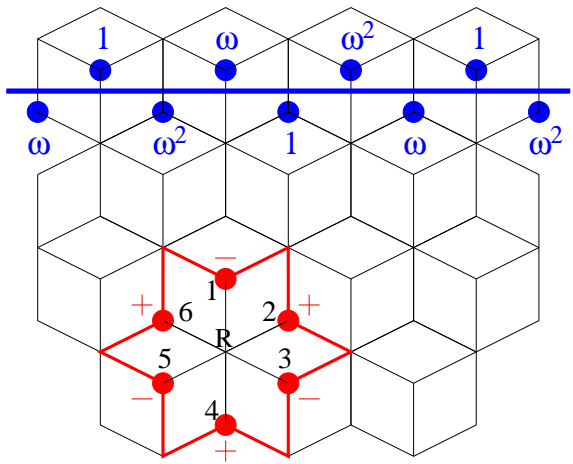}
	\caption{(Color online) Dice lattice localized eigenstate denoted by thick (red) closed loop, surrounding the site labeled by $R$. 
	One of two non-trivial loops is indicated by a thick straight (blue) line. All sites with 
	some particle weight on them are indicated by a filled circle.
	The amplitude is indicated on every site with non-zero weight.
	From this picture we can understand why the localized states are eigenstates, again 
	invoking the picture of destructive interference. Consider the sites marked 1 and 2.
	Hopping from these sites can occur to either ${\bf R}$ or the site right outside the 
	boundary of the localized state. in both cases, the hopping amplitude from sites 1 and 2 
	cancels out. Similar considerations for the other sites of the localized states yield
	the same result. For the non-contractible loop state, every site neighboring the sites with 
	nonzero weight have 3 hopping amplitudes contributing $1+\omega+\omega^2$. As long as this 
	sum vanishes, this is an exact eigenstate.}
	\label{fig:dice}
\end{figure}

The localized eigenstates of the dice lattice model are different than those of 
the kagome and pyrochlore lattices. Rather than residing in a loop around one or a number of plaquettes, the simplest
localized states here have nonzero weight on the 6 sites neighboring a central coordination-6 site with alternating 
signs, as illustrated in Fig.~\ref{fig:dice}. 

As for the kagome and pyrochlore lattices, a sum over all the localized states
surrounding every 6-coordinated site, can produce zero with
periodic boundary conditions
$A^{\dagger}_{{\bf q} = \pm (\frac{4\pi}{3},0)} = \sum_{\bf R} e^{+i {\bf q} \cdot {\bf R}} A^{\dagger}_{\bf R} = 0$.
Apart from these \emph{two} constraints, the localized states are all independent.
The number of coordination-6 sites on the lattice is the same as the number of unit cells, and so we have accounted
for $N-2$ states. As in the kagome and pyrochlore models, we will find eigenstates composed of non-contractible loops around 
the torus.

In Fig.~\ref{fig:dice} we show one of the two non-trivial loop states which exist for this model, with the weights being integer 
powers of the factor $\omega = e^{ \pm i \frac{4\pi}{3}}$, for a total of 4 non-contractible loop states. These are exact 
eigenstates provided that $1 + \omega + \omega^2 = 0$ is satisfied. Indeed, $\omega = e^{ \pm i \frac{4\pi}{3}}$ are the two 
solutions of this equation.

\subsection{Honeycomb lattice p-band model}
\label{sec:honeycomb-lattice-p}

Another hopping model with flat bands is the p-band hopping model on the honeycomb
lattice introduced in Ref.~\onlinecite{Wu:prl07}. 
In this model, only the planar $p_{x,y}$ orbital states are considered at each 
lattice site. It is convenient to describe any superposition of the two orbital states on a site with an orbital unit vector vector ${\vec p} = (p_x,p_y)$ representing the state
$ |{\vec p} \rangle = \cos{\phi} |p_x \rangle + \sin{\phi} |p_y \rangle $
with $\phi$ the angle between the arrow and the x-axis.
The hopping is assumed to occur only between orbital states with the 
orbital vector parallel to the link. The resultant tight binding Hamiltonian is
$
{\mathcal H} = t \sum_{\langle i j \rangle}
\left[ {\vec p}_i^{\dagger} \cdot \left( {\bf r}_i - {\bf r_j} \right) \right]
\left[ {\vec p}_j^{\phantom\dagger} \cdot \left( {\bf r}_i - {\bf r_j} \right) \right] 
+ h.c. 
$ 
where the nearest-neighbor distance is taken as unity,
and ${\vec p}_j^{\dagger}$ ( ${\vec p}_j$) are the creation (annihilation) operators 
for a particle at site $j$ in the orbital state $|{\vec p}\rangle$.

\begin{figure}
	\centering
		\includegraphics[width=3.0in]{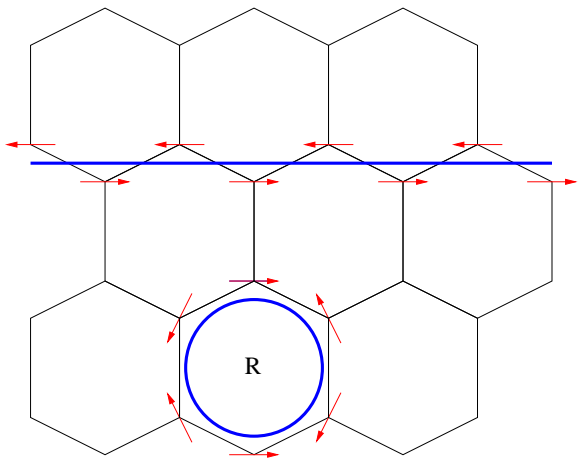}
	\caption{(Color online) p-band honeycomb local eigenstate and non-contractible loop state.
	The orbital states are denoted by a vector (in the present case the vector coordinates can be 
	chosen real). The orbital vectors are always perpendicular to one link emanating from the 
	site, and with the special form of hopping assumed in the model, the hopping amplitude on 
	this link vanishes in this state.}
	\label{fig:p_honeycomb}
\end{figure}

The two particle-hole symmetric flat bands in this model
touch dispersive bands at ${\bf q} = 0$. For the low-energy flat band, the local 
eigenstates are illustrated in Fig.~\ref{fig:p_honeycomb}
where the arrows denote the orbital state vectors.
The high-energy flat band states have all the orbital-arrows pointing in the 
opposite direction on \emph{one sublattice}, 
${\vec p} \rightarrow - {\vec p}$.

Exactly as for the kagome and pyrochlore lattice models, summing the localized states on all plaquettes
results in \emph{zero} $a^{\dagger}_{{\bf q} = 0} = \sum_{\bf R} a^{\dagger}_{\bf R} = 0$.
The localized states produce $N-1$ linearly independent states, leaving one state in each flat band unaccounted for.
Two non-contractible loop states exist (one on each handle of the torus), with the same energy as the flat band states, their details illustrated in 
Fig.~\ref{fig:p_honeycomb}. A superposition of all the translations of a non-contractible loop state, results in a 
${\bf q} = 0$ state. The two additional states then account for two ${\bf q} = 0$ modes in the band structure,
and explain why the dispersive bands must touch the flat bands at this momentum point.

\subsection{Dice lattice p-band model}\label{gapped1}
\label{sec:dice-lattice-p}

Another model that can be analyzed in a similar manner is the same p-band hopping introduced in Ref.~\onlinecite{Wu:prl07},
on the dice lattice. This model is different from the others presented here in that the flat bands are \emph{separated} 
from all the dispersive bands by finite energy gaps. We will use the same framework 
to understand why the flat bands are gapped in this case.

The band structure for this hopping model has two degenerate flat bands at $\epsilon = 0$,
and two pairs of particle-hole symmetric bands with energies
\begin{equation}
\epsilon({\bf q}) = 
\pm t \sqrt{2} \sqrt{(6+A) \pm \sqrt{2 (2 A^2 - 3 A -3 B)} }
\; ,
\end{equation}
with
\begin{equation}
\begin{split} &
A = \sum_{\mu = 1}^3 \cos{({\bf q} \cdot {\bf a}_{\mu})}
\\ &
B = \sum_{\mu = 1}^3 \cos{({\bf q} \cdot \left({\bf a}_{\mu} + 2 {\bf a}_{\mu + 1} \right) )}
\; ,
\end{split}
\end{equation}
where the three vectors ${\bf a}_{1,2,3}$ are the same minimal length Bravais lattice
vectors indicated in Fig.~\ref{fig:convention} (now with length $\sqrt{3}$), 
making $A,B$ two functions that are invariant under the full symmetry group of the model.
The dispersive bands are separated from the flat bands by a gap of $\Delta = \sqrt{2} t$.

\begin{figure}
	\centering
		\includegraphics[width=3.0in]{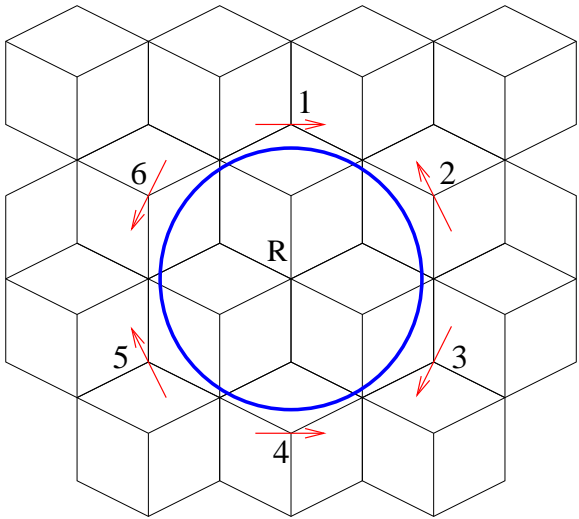}
	\caption{(Color online) Type I p-band dice lattice local eigenstate.}
	\label{fig:p_dice}
\end{figure}

\begin{figure}
	\centering
		\includegraphics[width=3.0in]{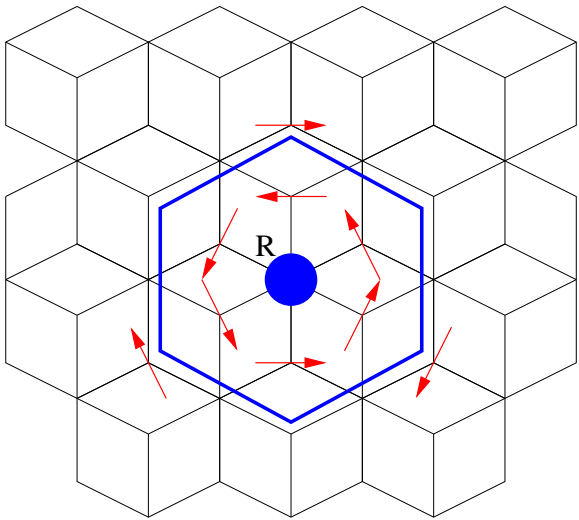}
	\caption{(Color online) Type II p-band dice lattice local eigenstate.}
	\label{fig:p_dice_2}
\end{figure}

Two local eigenstate modes can be found, on the same area unit surrounding a coordination-6 site.
They are illustrated in Fig.~\ref{fig:p_dice}  (type I) and Fig.~\ref{fig:p_dice_2} (type II),
using the same conventions we have introduced for the honeycomb p-band model in the previous subsection.
Note that the arrows indicating the orbital state are always in one of 6 discrete directions, 
with the angles $0^{\circ}, \pm 60^{\circ}, \pm 120^{\circ}, 180^{\circ}$ from the x-axis direction.
In these states, the arrow directions are always perpendicular to one link emanating from the site.

Naively, all these states (type I and II) on different area units are linearly independent, resulting in $2N$ states 
that exhaust the number of states in the flat bands.
However, as with the previous models explored in this paper, we will find a number of constraints that show 
that this is not the case. 

\begin{figure}
	\centering
		\includegraphics[width=3.0in]{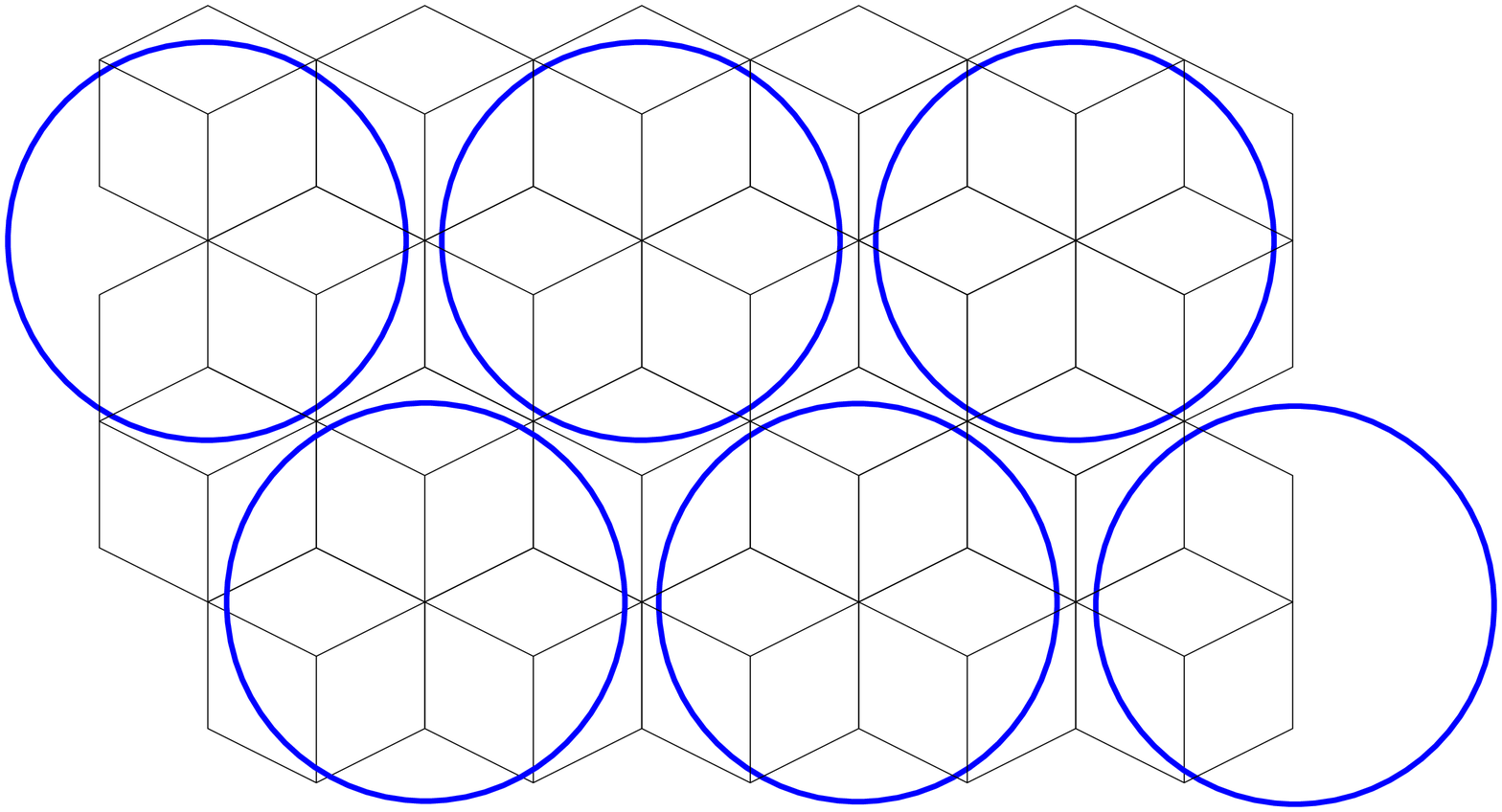}
	\caption{(Color online) Non-overlapping set of area units. Summing over type I states on these area units adds up to zero.}
	\label{fig:dice_3}
\end{figure}

Summing over the type I states on a set of non-overlapping area units that cover the entire plane results in zero,
when we have periodic boundary conditions. An
appropriate choice of the area units to sum over is illustrated in Fig.~\ref{fig:dice_3}.
However, there are 4 such distinct sets covering the entire plane, and so there are 4 different sums (involving 
different sets of states) giving zero. The 4 sets of type I states are related by the Bravais translations of the lattice.

\begin{figure}
	\centering
		\includegraphics[width=3.0in]{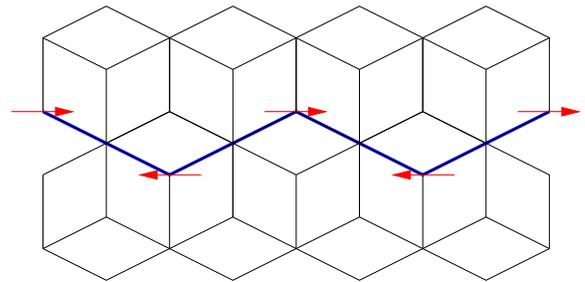}
	\caption{(Color online) Non-contractible loop states. Only those sites with arrows on them have non-zero occupation.}
	\label{fig:dice_4}
\end{figure}

As in the other models we have discussed, non-contractible loop eigenstates can be constructed, as illustrated
in Fig.~\ref{fig:dice_4}. The non-contractible loop eigenstates always come in pairs - one around each
non-trivial loop on the torus. However, there are 4 pairs of distinct such states related by translation. The quadrupling
is closely related to the 4 different unit area sets covering the plane discussed in the previous paragraph.
The non-contractible loop states we are presenting here always lie on the edges of one of the 4 sets of area units.

\begin{figure}
	\centering
		\includegraphics[width=3.0in]{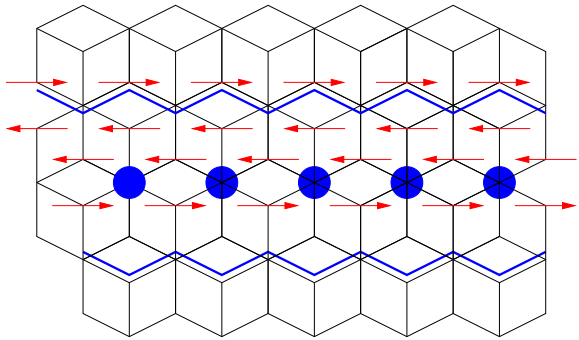}
	\caption{(Color online) Sum of type II states (marked by filled circles at their centers) on a non-contractible strip of area units. Only those sites with arrows on them have non-zero occupation.}
	\label{fig:dice_6}
\end{figure}

There are 4 additional constraints involving sums over type II localized eigenstates, and the non-contractible loop states.
Adding the type II states on every area unit on an infinite length strip, comprising a non-contractible loop around the torus,
we find the resultant state depicted in Fig.~\ref{fig:dice_6}. The resultant state is a sum of 4 non-contractible loop
states, of the kind depicted in Fig.~\ref{fig:dice_4}, and so the type II states and the non-contractible loops are not linearly
independent. The sum of type II on a strip translated by one
Bravais lattice vector, perpendicular to the strip gives another independent relation between the type II states and the 
non-contractible loop states.
Finally, for strips around the other handle on the torus, we find two additional such relations. 

In total, we have $2N$ localized eigenstates, 8 non-contractible loop states, and $4+4=8$
vanishing linear combinations of these $2N+8$ states. We therefore have precisely $2N$ linearly independent
states, all with the same energy of the flat bands, exhausting the number of states in the flat bands. 
Since no extra states at this energy need to be accounted for by the dispersive bands, a gap can occur in 
principle, and indeed shows up in practice.

\section{Discussion}
\label{discussion}

\subsection{summary}
\label{sec:summary}

We have found that the presence or absence of band touchings can be
understood for a variety of frustrated hopping models by a careful
counting of linearly independent localized states.  Crucial to this
counting is the presence, which we found for all the models examined
here, of non-contractible loop states.  In all cases, a vanishing
superposition of the localized states was found, always involving a sum
over area patches that cover the entire lattice, but some with different
weights in the summation ( non-zero ${\bf q}$ for instance).  When
working in a toroidal geometry, for which momentum ${\bf q}$ remains a
good quantum number, the missing states eliminated by this vanishing
superposition were recovered as non-contractible loop states.  The
counting, for the case of a single flat band, is as follows.  The total
number of independent states with the flat band energy is $N - M + d L$,
where $N$ is the number of unit cells, $M$ is the number of localized
state independent superpositions that vanish (the number of missing
states), and $L$ is the number of different `flavors' of
non-contractible loop states for each of the $d$ handles of the torus
(in $d$ dimensions). When this is larger than $N$, the flat band is
degenerate with some other band, at a finite number of points in
momentum space.

\subsection{Are frustrated hopping models frustrated?}
\label{sec:are-frustr-hopp}

We have referred to the hopping models discussed in this paper as
``frustrated'', simply because (apart from the p-orbital honeycomb
example), they reside on lattices exhibiting strong geometrical
frustration for antiferromagnetism.  It is interesting to see if this
abuse of terminology has any truth to it. Consider a general tight
binding Hamiltonian with a flat band
\begin{equation}\label{hoppingH}
{\mathcal H} = \sum_{i j} c_i^{\dagger} t_{i j} c_j
\; ,
\end{equation} 
where the indices $i,j$ include all the generalized coordinates of the
particles (position, orbital state etc...). 
Choosing the particles to be fermionic, adding a spin $\frac{1}{2}$ index and an on-site interaction term, we have
\begin{equation}\label{generalH}
{\mathcal H} = \sum_{i j \alpha} c_{i \alpha}^{\dagger} t_{i j} c_{j \alpha}
+ U \sum_j c_{j \uparrow}^{\dagger} c_{j \uparrow} c_{j \downarrow}^{\dagger} c_{j \downarrow}
\; .
\end{equation}
Note that the interaction corresponds to the physical on-site repulsion only if 
the indices $j$ represent only spatial coordinates, but this is only a minor 
complication we will ignore in this short and simplistic analysis.

At half filling of the fermions, for infinite repulsive $U$, we get a
Mott insulating phase, with one fermion occupying each state of the
(weighted) network of sites $j$ described by the matrix $t_{i j}$.  For
finite but very large $U$, the virtual hopping of particles results in a
Heisenberg model splitting the energy of the insulating states. It is
very easy to show that the general Heisenberg model that emerges is
\begin{equation}
{\mathcal H} = \sum_{i j} J_{i j} {\bf S}_i \cdot {\bf S}_j + {\mathcal O}\left(
\frac{t^3}{U^2} \right)
\; ,\label{eq:3}
\end{equation} 
with $J_{i j} = \frac{4 t_{i j}^2}{U}$, where ${\bf S}_j$ are the spin
$\frac{1}{2}$ operators.  The matrix structure that is responsible for
the flat bands carries over to the exchange interaction matrix. In
particular, if the hopping matrix takes on only two values, $0$ and $t$
then $J_{i j} = \frac{4 t}{U} t_{i j}$ and the matrices simply differ by
a multiplicative factor.  

When $t_{ij}$ is non-zero for only nearest neighbor sites $i,j$, the
spin model ``descending'' from the hopping Hamiltonian is indeed
geometrically frustrated.  It is interesting to consider, however, a
possibly more direct connection between flat bands and frustration.
Indeed, somewhat rough arguments suggest that, fairly generally, the
classical spin Hamiltonian with exchange $J_{i j} = \frac{4 t}{U} t_{i
  j}$ has an extensive ground state degeneracy when the lowest energy
eigenstates of $t_{ij}$ form a flat band.  We discuss some additional
conditions below.

The connection between flat bands and ground state degeneracy is through
the Luttinger-Tisza method for finding ground states of classical spin
models.  The idea is the following.  We first trade the normalization
constraint on the spins, $|{\bf S}_j|=S$, for the weaker condition
\begin{equation}
  \label{eq:4}
  \sum_j |{\bf S}_j|^2=N_s S^2,
\end{equation}
where $N_s$ is the number of spins.  With this weaker constraint, it is
simple to minimize ${\mathcal H}$ in Eq.~(\ref{eq:3}).  This is
guaranteed to give an energy which is at least as low as for the minima of
${\mathcal H}$ taking proper spin normalization into account. The general
solution is an arbitrary {\sl real} linear combination of the minimum energy
eigenstates of $J_{i,j}$, and hence of $t_{i,j}$.  Because the hopping
matrix is real, the eigenstates may also be chosen real.  When there are
$n_f$ flat bands, there are a total of $n_f N$ such eigenstates with
minimum energy, for a lattice containing $N$ unit cells.   The solution is
\begin{equation}
  \label{eq:5}
  {\bf S}_j = \sum_{a=1}^{N n_f} {\bf s}_a \phi_j(a),
\end{equation}
where $\phi_j(a)$ is the $a^{th}$ eigenstate of $t_{ij}$, and the ${\bf
  s}_a$ is an unknown vector of real coefficients for each $a$.  The
total number of variables that may be varied is then $3$ real numbers
for each $a$, and hence $3N n_f$ real numbers.  Now we can attempt to
impose the necessary constraints to get a physical minimum for the
classical spin model.  These consist of one constraint per spin on the
spin magnitude.  If the lattice contains $n_b$ spins in its basis
(i.e. sites in the unit cell), then this gives $N n_b$ constraints.
Subtracting the number of constraints from the number of variables gives
a total of $(3n_f-n_b)N$ degrees of freedom remaining for physical
minima of the classical spin model.  Thus, for $n_b< 3n_f$, we are led
to expect an extensive degeneracy (macroscopic entropy) of spin ground
states.  This counting is certainly crude, and since the normalization
constraints are non-linear, not entirely rigorous.  For instance, if
$n_b = 3 n_f$, it is probably the case that extensive ground state
entropy may or may not be present, depending upon other details of the
model.  This is born out, for instance, by the case of the nearest
neighbor kagome model, for which there is indeed an extensive ground
state entropy, while $n_b=3,n_f=1$. However, we believe the conclusion
provides a reasonable qualitative guide, though the estimate of the
entropy density is probably unreliable.  Thus a large $U$ Hubbard model
with a kinetic energy with flat minimum energy bands indeed, when
$n_b<3n_f$, exhibits macroscopic ground state entropy, the classic
signature of frustration.  

\subsection{For the future}
\label{sec:future}

The discussion in this paper is only a prelude to the study of
interacting bosons and fermions in flat band systems.  As discussed in
the introduction, this is understood in generality only for low density,
below the appropriate ``close packing'' threshold.  In this case the
ground states are completely localized and minimize simultaneously the
kinetic and interaction energies.  Above this close packing density, the
problem becomes much more intricate, and it is reasonable to expect
delocalized ``liquid'' ground states.  How to attack the problem in the
range of densities for which the particles can still be accommodated
in the flat band(s) but above close packing is an interesting open
problem.  It is intriguing to speculate that liquid states in this
regime, at least for weak interactions, may have unconventional
properties.  With the increasing accessibility of such Hamiltonians in ultra-cold atomic
systems in optical lattices, clarification of this regime may well come
experimentally rather than theoretically.

\begin{acknowledgments} 
 
The authors would like to acknowledge Congjun Wu and Sankar Das-Sarma for
early collaboration on related work, and John Chalker, for   
illuminating discussions. 
L.B. was supported by the Packard Foundation and the National
Science Foundation through \mbox{grant DMR04-57440}.
  
\end{acknowledgments}

\appendix
\section{Another kagome model}
\label{appendix1}

\begin{figure}
	\centering
		\includegraphics[width=1.5in]{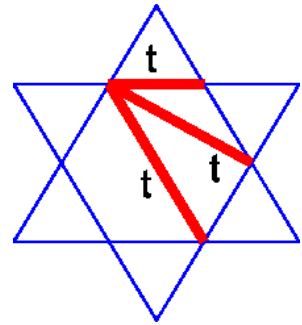}
	\caption{(Color online) Tight binding model with equal hopping amplitudes on three different link types.}
	\label{fig:BFG}
\end{figure}

Having identified a connection between frustrated magnetic models
and flat bands, we examine one additional model - one inspired by 
Ref.~\onlinecite{Balents:prb02}, and mirroring its geometric structure.
The magnetic model in Ref.~\onlinecite{Balents:prb02} has been shown to support a 
spin liquid ground state. The magnetic model Hamiltonian consists of exchange 
interactions of equal strength on 3 different link types. This is described by 
an exchange matrix. We take this same matrix structure and construct a simple tight 
binding model with it. The Hamiltonian is precisely of the form of Eq.~\eqref{hoppingH}
with the indices corresponding to the lattice sites of the kagome lattice.
The hopping amplitudes are illustrated in Fig.~\ref{fig:BFG},
and we will refer to this hopping model as the kagome-3 model.

Our analysis finds the band structure of this model consists of \emph{two} 
degenerate flat bands at energy $\epsilon_0 = +2 t$ and one dispersive band 
$\epsilon_1 = -t \left(4 + 2 \Lambda({\bf q}) \right)$ with $\Lambda({\bf q})$ the 
\emph{same} function introduced in the introduction.
The single dispersive band has a minimum energy of $-10 t$ at ${\bf q} = 0$,
and a maximum energy of $-t$ at wavevector ${\bf q} = \frac{4 \pi}{3} {\hat x}$
in our conventions. The flat bands are therefore \emph{gapped} from the dispersive 
band with a gap of $\Delta = 3t$.
The convention we use for the 3 vectors illustrated in 
Fig.~\ref{fig:convention}, is $a_1 = {\hat x}$, 
$a_2 = -\frac{1}{2}{\hat x} + \frac{\sqrt{3}}{2}{\hat y}$, and 
$a_3 = -\frac{1}{2}{\hat x} - \frac{\sqrt{3}}{2}{\hat y}$.

\begin{figure}
	\centering
		\includegraphics[width=3.0in]{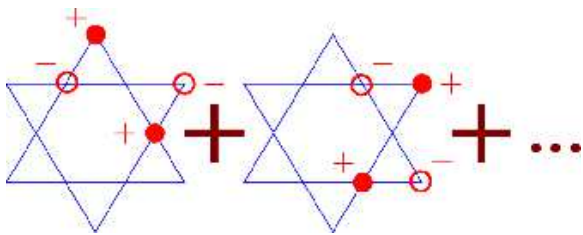}
	\caption{(Color online) Illustration of the local constraint in the kagome-3 hopping model.
	Localized eigenstates living on the bowtie plaquettes are summed over the 6 plaquettes bordering a hexagonal plaquette in the kagome lattice. The figure shows two of these localized states, with the correct relative phase needed for the summation to vanish. Only those sites with non-zero weight are denoted by (red) circles (filled for $+1$, unfilled for $-1$). }
	\label{fig:local}
\end{figure}

\begin{figure}
	\centering
		\includegraphics[width=3.0in]{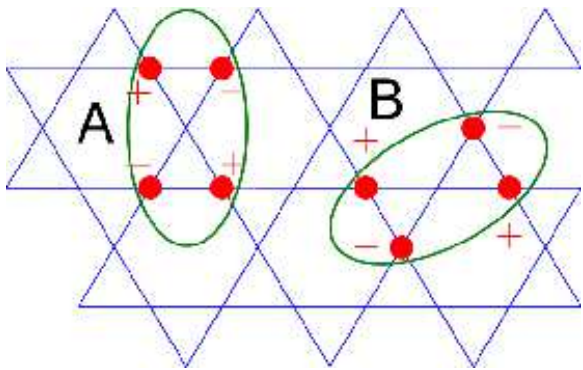}
	\caption{(Color online) Localized eigenstates of the kagome-3 hopping model.
	The states have non-zero weight only on the 4 sites surrounding a `bowtie' plaquette
	(the encircled regions). The sites with nonzero weight are marked by filled (red) 
	circles, and with their relative signs indicated next to them. There are a number 
	of different `flavors' of bowtie plaquettes (and localized states) - 3 per unit cell.
	However there are only 2 independent bowtie states per unit cell. Here we show one choice
	of two independent bowtie states marked as $A$ and $B$.}
	\label{fig:kagome_BFG2}
\end{figure}

Proceeding as for the other models in this paper, the localized states we find are illustrated in Fig.~\ref{fig:kagome_BFG2}, and live on ``bowtie'' plaquettes.
As in the pyrochlore lattice, naively there are 3 
different ``flavors'' of bowtie plaquettes (and localized states).
These can most easily be identified by considering how many bowties states involve a single up-pointing triangle (corresponding to a unit cell). Were these states all linearly 
independent, we would have $3 N$ states in the flat band, rather than $2 N$
(here as before, $N$ is the number of unit cells). However, taking a
bowtie plaquette state, rotating it around a hexagonal plaquette to produce 6 bowtie 
plaquette states around the hexagon ( see fig.~\ref{fig:local}), and finally summing these 6 wavefunctions, produces zero. For each hexagonal plaquette, of which there are $N$, there is one such constraint, and we find there are only 2 flavors of independent bowtie states -
marked $A,B$ in Fig.~\ref{fig:kagome_BFG2}.

\begin{figure}
	\centering
		\includegraphics[width=3.0in]{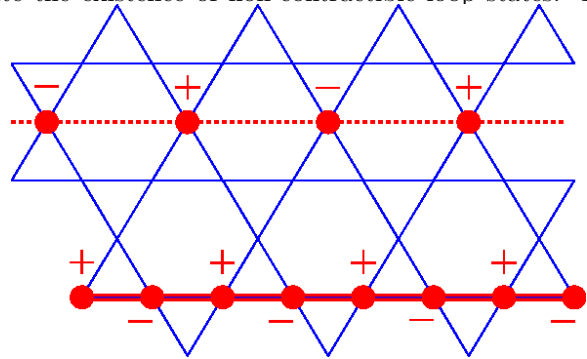}
	\caption{(Color online) Non-contractible loop states of the kagome-3 model. There are two types of the 
	non-contractible loops. One kind are the exact same states appearing in the 
	kagome model, and is denoted by a continuous (red) path between the sites with nonzero weight.
	The other kind of non-contractible loop states are denoted by dashed line, and consist
	of 2 sites with nonzero weight and alternating sign on each hexagonal plaquette in a chain 
	of plaquettes.}
	\label{fig:loopstates}
\end{figure}

As in all the other examples we give in this manuscript, there are additional 
non-local constraints, which mandate the existence of non-contractible loop states.
The constraints we find involve vanishing summations over the bowtie plaquette states of 
type A ( see Fig.~\ref{fig:kagome_BFG2} ), 
with wavevectors ${\bf q} = \frac{1}{2} {\bf b}_2, \frac{1}{2} \left( {\bf b}_1 + {\bf b}_2 \right)$ (but \emph{not} ${\bf q} = \frac{1}{2} {\bf b}_1$).
In the exact same manner,
the sums over the type B plaquette states, with wavevectors 
${\bf q} = \frac{1}{2} {\bf b}_1, \frac{1}{2} \left( {\bf b}_1 + {\bf b}_2 \right)$
(but \emph{not} ${\bf q} = \frac{1}{2} {\bf b}_2$), also vanish.
We therefore have 4 constraints, and since there are a total of $2N$
flat band states, we expect 4 non-contractible loop states to exist.
These can easily be found graphically, and are illustrated in Fig.~\ref{fig:loopstates}.

As in section~\ref{gapped1}, we find that in order for gapped flat bands to appear,
we needed two different localized eigenstates occupying the same area unit, as well as 
non-contractible loop states that are not associated with one of the two sets of local eigenstates.

\end{document}